\documentclass{appolb}
\usepackage{epsfig}

\newcommand{\zp}[1]{Z. Phys. {\bf #1}}

\begin{document}

\title{ARE NEUTRINOS DIRAC OR MAJORANA PARTICLES?\thanks{Presented by M. Zra\l ek 
at the XXIII School of Theoretical Physics, Ustro\'n'99,
Poland, September 15-22, 1999.}$\;$\thanks{Work supported in part by the
Polish Committee for Scientific Research under 
Grants Nos. 2P03B08414 and 2P03B04215. 
J.G. would like to thank also the Alexander von Humboldt-Stiftung for 
fellowship.}}
\author{M. CZAKON AND M. ZRA\L EK
\address {Department of Field Theory and Particle Physics, Institute 
of Physics, \\ 
University of
Silesia, Uniwersytecka 4, PL-40-007 Katowice, Poland} \\ \vspace{.5cm}
J. GLUZA
\address {Department of Field Theory and Particle Physics, Institute 
of Physics, \\ University of
Silesia, Uniwersytecka 4, PL-40-007 Katowice, Poland, \\
DESY Zeuthen, Platanenallee 6, 15738 Zeuthen, Germany}
}
%\date{}
\maketitle

\begin{abstract}
In spite of the general belief that neutrinos are Majorana particles, their
character should be revealed experimentally. We begin by  discussing why it is so
difficult in terrestrial experiments. If
neutrinos are Majorana particles, the first signal should come from
neutrinoless double $\beta $ decay. Still the search for such a  decay of
various nuclei is negative. We outline how the present knowledge
of neutrino masses and mixing matrix elements combined with the bound
from $\left( \beta \beta \right) _{0\nu }$ decay could help to determine
their nature.
\end{abstract}
\section{Introduction}

There are two main problems in neutrino physics. First is the problem of
neutrino masses, which in the light of present data \cite{sup} seems to be solved.
Neutrinos are massive. The second is that of the neutrino nature.
As massive they can be Dirac (with particles and
antiparticles being different objects: $\nu \neq \bar{\nu}$) or Majorana (with 
particles and antiparticles being the same, just as for photons: $\nu = \bar{\nu}$).
An experimental distinction between these two seems to be much more complicated
than the confirmation of non-vanishing mass. While experimentalists are trying 
to find some way of doing it, theorists have no doubts. They widely believe
in the Majorana nature \cite{maj}. Almost all extensions of the Standard Model (SM) predict it.
The only way to have Dirac neutrinos is to impose lepton number conservation. However,
there is no particular reason for this, since it is not a fundamental quantity like the
electric charge. If we do, we immediately run into trouble. 
Let us mention only a loss of the natural `see-saw' mechanism 
to
explain the smallness of the neutrino mass. 

Theoretical reasons aside, the scientific method obliges us to perform experimental studies,
that would falsify either option. So then, why is it difficult? Naively one might think it to be
rather easy. Imagine for example, neutrinos from  $\pi ^{+}$ decay 
($\pi ^{+}\rightarrow \mu ^{+}\nu _\mu $) scattering on a nuclear target. The result is a
flux of $\mu ^{-}$ (antineutrinos $\overline{\nu }_\mu $ coming from $\pi ^{-}$ decay
($\pi ^{-}\rightarrow \mu ^{-}\overline{\nu }_\mu )$ always produce
antiparticles $\mu ^{+}$).
Unfortunately the lepton number L (L($\nu _\mu $)= +1, L($\nu _\mu $)=$ -1$) 
is
not the only property characterizing neutrinos. We know also from
experiment \cite{exp} that neutrinos and antineutrinos have opposite helicity ($\nu
_\mu =\nu (-)$ and $\overline{\nu }_\mu =\nu (+)$). Therefore, we are not
able to state which is responsible for $\mu ^{-}$ ($\mu ^{+}$)
production, lepton number conservation or helicity. In the first case the
left-handed neutrino fields $\nu _L(x)$
\begin{equation}
\nu _L(x)=\int \frac{d^3k}{\left( 2\pi \right) ^3}\left(
A(-)e^{-ikx}-B^{+}\left( +\right) e^{ikx} \right) \chi (-)
\end{equation}
are composed of two different operators (see \cite{zr} for a detailed definition).
$A(-)$ which annihilates particles has  negative helicity and $B^{+}(+)$ which
creates particles has positive helicity. 

For a massless Majorana field $N_L(x)$ only one operator $A=B\equiv a$ appears
\begin{equation}
N_L(x)=\int \frac{d^3k}{\left( 2\pi \right) ^3}\left(
a(-)e^{-ikx}-a^{+}\left( +\right) e^{ikx} \right) \chi (-) . 
\end{equation}
In order to check whether lepton number conservation $(A\neq B)$ or particle
helicity $(a(-)\neq a(+))$ is responsible for $\mu ^{-}$($\mu ^{+}$)
production, we have to compare neutrino interactions in the same helicity
states
\begin{equation}
A(-)\;\;\; {\rm  with }\;\;\;B(-) ,
\end{equation}
or
\begin{equation}
A(+) \;\;\; {\rm with } \;\;\; B(+) .
\end{equation}
Unfortunately, the visible neutrino interactions are such that only
particles in the states A(-) and B(+) are produced. No neutrinos in the states
$A(+)$ and $B(-)$ appear in known experiments.

In the next section we would like to show the connection between the presence
(or absence) of the states given in Eqs. (3) and (4) with the symmetries of the
theory.

Next in Chapter 3 examples which explain the origin of the experimental difficulties
of discerning Dirac from Majorana neutrinos are given.
The main background being the  small mass of neutrinos which causes that they are produced as highly
relativistic particles and their visible left-handed interaction.

It is common belief that the first place to search 
 is the neutrinoless double $\beta$ decay $%
((\beta \beta )_{0\nu })$ of nuclei. Unfortunately up to now such a decay has
not been found and experimental data gives lower bounds on $%
(\beta \beta )_{0\nu }$ decay modes of various nuclei. These in turn lead to 
the limit \cite{eff} on the so-called effective neutrino mass $\left\langle m_\nu \right\rangle $
\begin{equation}
\left| \left\langle m_\nu \right\rangle \right| \equiv \left| \sum
U_{ei}^2m_i\right| <0.2\; {\rm eV}. 
\end{equation}
There are plans to increase the sensitivity of the bound(s) down to 0.01 or
even 0.001 eV \cite{pla}. If $\left\langle m_\nu \right\rangle \neq 0(=0)$ the
neutrinos are massive Majorana (Dirac) particles. Currently, however, the bound (5) alone is
not conclusive. There are nevertheless different experiments from which
independent information on the neutrino mixing matrix elements $U_{ei}$ and
masses $m_{i{ \rm  }}$ can be inferred. Then, we can check whether the bound
(5) is satisfied or not. If not, neutrinos are Dirac particles. If it is
satisfied, no conclusion can be drawn. Such an analysis
is performed in Chapter (4). Finally, in Chapter (5) the conclusions are given.

\section{Dirac or Majorana nature of particles, and symmetries.}

We would like to explain how the particle content of a theory is connected with
its symmetries.

We believe up to now \cite{bie}, that the fundamental symmetry of any theory which
describes elementary particle interactions is Lorentz invariance. This
statement means precisely that the theory must be invariant under the proper
orthochronous group of Lorentz transformations $L_{+}^{\uparrow }$. For massive
particles, they  mix states with all helicities, for massless, 
helicity is Lorentz invariant. So, from $L_{+}^{\uparrow }$
invariance it follows that:
\begin{itemize}
\item  for massive particles $(m\neq 0)$ with spin j all states
\end{itemize}
\begin{equation}
\left| \overrightarrow{p},\lambda \right\rangle \;\; { \rm  for }\;\; \lambda
=-j,-j+1...,+j 
\end{equation}
must be present in the theory
\begin{itemize}
\item  for massless spin j particles $(m=0)$ only one state
\end{itemize}
\begin{equation}
\left| \overrightarrow{p},\lambda =j\right\rangle { \rm  or }\left| 
\overrightarrow{p},\lambda =-j\right\rangle
\end{equation}
must be introduced.

For example, it is possible to built a theory which has $L_{+}^{\uparrow }$
invariance with three helicity states of the $W^{+}$  $\lambda
=-1,0,+1$ with no $W^-$ and a photon of one  polarization e.g$.\left| photon,\lambda
=+1\right\rangle $ or a neutrino in the state $\left| neutrino,\lambda
=-1/2\right\rangle $.

The next symmetry is invariance under the CPT transformation \cite{cpt} which changes
particles into antiparticles and helicity $\lambda \rightarrow -\lambda .$

\begin{equation}
CPT\left| \overrightarrow{p},\lambda \right\rangle _{particle}=\left| 
\overrightarrow{p},-\lambda \right\rangle _{antiparticle.} 
\end{equation}
In any theory with CPT symmetry,  particles and antiparticles with
opposite helicities must exist.
In our example this means that $W^{-}$ particles with $\lambda =\pm 1.0$, and an
antiphoton with $\lambda =-1$, and antineutrinos with $\lambda =+1/2$ must be
present.

There are theories like QED where also the separate symmetries C, P and T hold.
The helicity states transform as 
\begin{equation}
P\left| \overrightarrow{p},\lambda \right\rangle =\eta _Pe^{i\pi \lambda
}\left| -\overrightarrow{p},-\lambda \right\rangle ,
\end{equation}
\begin{equation}
T\left| \overrightarrow{p},\lambda \right\rangle =\eta _Te^{i\pi \lambda
}\left| -\overrightarrow{p},\lambda \right\rangle ,
\end{equation}
and 
\begin{equation}
C\left| \overrightarrow{p},\lambda \right\rangle _{particle}=\eta _Ce^{i\pi
\lambda }\left| \overrightarrow{p},\lambda \right\rangle _{antiparticle} . 
\end{equation}
For massive particles these symmetries do not introduce new necessary
particle states above those already present because of Lorentz invariance
and CPT symmetry.
For massless particles, however, P leads to the existence of particle
(antiparticle) states with opposite helicities.
Once more in our example there has to be a photon and an antiphoton.
\begin{figure}
\epsfig{figure=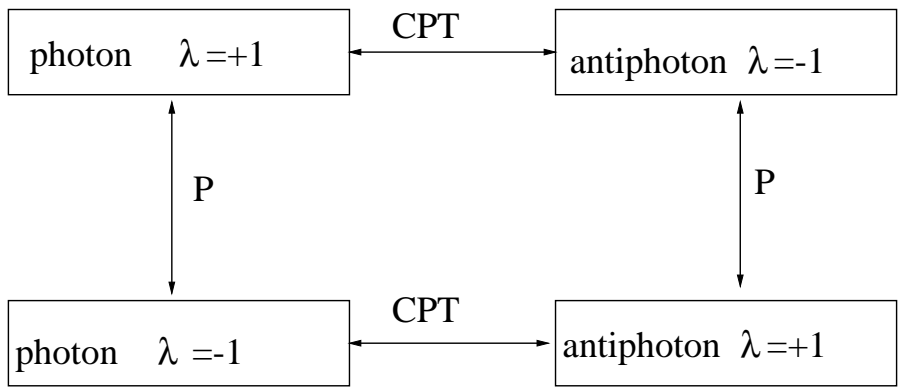, height=2 in}
%angle=270 hscale=50 vscale=50 hoffset=50voffset=-325}
\caption{Four `photon' states connected by CPT and P transformations.}
\end{figure}

Now we can go back to our previous statement: in order to determine the
nature of neutral objects we need to compare the interaction of particles and
antiparticles in the same helicity states.
\begin{equation}
\left| \overrightarrow{p},\lambda \right\rangle _{particle}\;\;\;
{\rm  with }%
\;\;\; \left| \overrightarrow{p},\lambda \right\rangle _{antiparticle}. 
\end{equation}
In a theory with C, P and T symmetry:

(i) such states exist for massive and massless particles

(ii) from C symmetry particles and antiparticles interact in the same way so
there is no way to distinguish them.

This means that in those fully symmetric theories, there are only MAJORANA neutral particles.
That is why photons must be Majorana particles in QED.

All looks different in theories where C, P, T symmetries do not hold
(like in the weak interactions).

For massive particles two states (Eq. (12) exist and we can compare their
interactions. Particles and antiparticles in the same helicity states
can interact
(i) in different ways or (ii) identically.

In case (i):

\begin{itemize}
\item  some additive quantum number exists, which differentiate particles from
antiparticles,

\item  particles and antiparticles are not the same,

\item  it is the case of massive Dirac neutrinos (described by bispinors) with lepton number
conservation.
\begin{equation}
\left( \begin{array}{c} \nu _R \\ \nu _L \end{array} \right) \equiv \Psi _D\neq \Psi _D^C. 
\end{equation}
In the case (ii):

\item  additive quantum numbers cannot exist,

\item  particles and antiparticles are not indistinguishable, they are
Majorana objects,

\item  there are two important examples of such particles:
the $Z_0$ gauge boson, and massive Majorana neutrinos described by Majorana bispinors 
\begin{equation}
\left( \begin{array}{c} \nu _R \\ \nu _L \end{array} \right) \equiv \Psi _M= \Psi _M^C. 
\end{equation}
For massless particles the symmetries do not require the existence of both
states in Eq. (12). It is possible to built theories where particles and
antiparticles in the same helicity state (i) do not exist or (ii) are
introduced.

In the same case (i):

\item  the discussion about Dirac or Majorana nature of such particles is
meaningless, there is nothing to compare,

\item  in the case of spin 1/2 objects there is a kinematical theorem \cite{pa},
which proves that Weyl neutrinos are identical with massless Majorana
neutrinos.

In the case (ii):

\item  two spinors $\nu _L$ and $\nu _R$ are introduced. As in the L-R
symmetric model, four states described by $A$($\pm )$ and $B(\pm )$
annihilation operators exist,

\item  objects $A$($\pm )$ and $B(\pm )$ can interact in different ways so we
have massless Dirac neutrinos (or if CP is conserved, two Majorana
neutrinos with opposite $\eta _{CP}$ parities)

\item  objects $A$($\pm )$ and $B(\pm )$ interact in the same way and we
have two identical massless Majorana neutrinos (these
Majorana neutrinos have the same $\eta _{CP}$ parity).
\end{itemize}

\section{Why is it difficult to distinguish experimentally Dirac and Majorana
neutrinos?}

There are two main reasons, which cause that practically it is impossible, at
least with the present experimental precision to determine nature of
neutrinos \cite{pra}. Firstly, the created neutrinos are usually relativistic
$(E>>m)$. On the other hand, cross sections for neutrino
interaction are proportional to the energy E, so that nonrelativistic neutrinos
interact with matter very weakly. Secondly, visible neutrino interactions are
either left-handed $\frac 12\gamma ^\mu (1-\gamma _5)$ for gauge bosons or
proportional to neutrino mass for scalar particles ($m_\nu /m_W$ for Higgs
particles).

The forthcoming examples will demonstrate these problems.

Let us assume that a beam of muon neutrinos with helicity $h_\nu$, scatters on a nuclear
target. To be more general we consider the neutrino charged current
interaction to be of  the form
\begin{equation}
L_{CC}=\frac g{\sqrt{2}}\left[ A_L\left( \overline{N} \gamma ^\mu
P_Ll\right) +A_R\left( \overline{N} \gamma ^\mu P_Rl\right) \right] W_\mu
^{+}+h.c. 
\end{equation}
with a left-handed $(A_L)$ and a hypothetical right-handed $(A_R)$ part.
Dirac neutrinos generate only $\mu ^{-}$'s (with helicity $h_\mu $)
\begin{equation}
\nu ^D+N\rightarrow \mu ^{-}+X ,
\end{equation}
with the amplitude proportional to 
\begin{eqnarray}
A_{\mu ^{-}}^D(h_\nu ,h_\mu ) & \sim & 
 A_L^{*}\left[ \left( E_\mu -2h_\mu p_\mu \right) 
\left( E_\nu -2h_\nu p_\nu \right) \right] ^{1/2} \nonumber \\
&+& A_R^{*} \left[
\left( E_\mu +2h_\mu p_\mu \right) \left( E_\nu +2h_\nu p_\nu \right)
\right] ^{1/2} 
\end{eqnarray}
where $E_\mu ,p_\mu $ ($E_\nu, p_\nu )$ is the  energy and momentum of the muons
(neutrinos).

Majorana neutrinos generate $\mu ^{-}$'s with exactly the same amplitude Eq.
(17) and $\mu ^{+}$'s. The amplitude for $\mu ^{+}$ production is now
proportional to
\begin{eqnarray}
A_{\mu ^{+}}^M(h_\nu ,h_\mu ) & \sim & A_L\left[ \left( E_\mu +2h_\mu p_\mu
\right) \left( E_\nu +2h_\nu p_\nu \right) \right] ^{1/2} \nonumber \\
&+& A_R\left[
\left( E_\mu -2h_\mu p_\mu \right) \left( E_\nu -2h_\nu p_\nu \right)
\right] ^{1/2}. 
\end{eqnarray}
In the laboratory frame we are able to obtain a  beam of muon neutrinos with helicity $
h_\nu =-1/2$ (e.g. from $\pi ^{+}\rightarrow \mu ^{+}\nu _\mu ).$
The cross section for $\mu ^{+}$ production is unfortunately proportional to 
\begin{eqnarray}
\sigma _{\mu ^{+}}^M(h_\nu =-1/2) & \sim & \left| A_L\sqrt{E_\nu -p_\nu }+\zeta
(h_\mu )A_R\sqrt{E_\nu +p_\nu }\right| ^2 \nonumber \\
& \approx & \left| A_L\frac{m_\nu }{%
\sqrt{2E_\nu }} +\zeta (h_\mu )A_R\sqrt{2E_\nu }\right| ^2. 
\end{eqnarray}
Both terms in (19) are small in the high $\beta$ limit.

For neutral current interactions the situation seems at first sight to be even more
promising. There are two characteristic features, which are
completely different for Dirac and Majorana neutrinos.

(i) the vector current $\overline{\nu }_M\gamma ^\mu \nu _M=0$, for Majorana neutrinos \\
and

(ii) Majorana neutrinos, as identical particles, need symmetrization.

Let us consider shortly both of them. The respective neutral current interactions are of the form
\begin{equation}
L_{NC}(D)=\overline{\nu }_D\gamma ^\mu \left( g_V ^D-g_A^D\gamma _5\right)
\nu _DZ_\mu , 
\end{equation}
and
\begin{equation}
L_{NC}(M)=\overline{\nu }_M\gamma ^\mu \left( -g_A ^M \gamma
_5\right) \nu _MZ_\mu . 
\end{equation}
Despite this striking difference, both cases are again indistinguishable \cite{pra}. Let us consider the
measurement of the total cross section for inclusive production (Fig.2)
\begin{equation}
\nu +N\rightarrow \nu +X. 
\end{equation}

\begin{figure}
\epsfig{figure=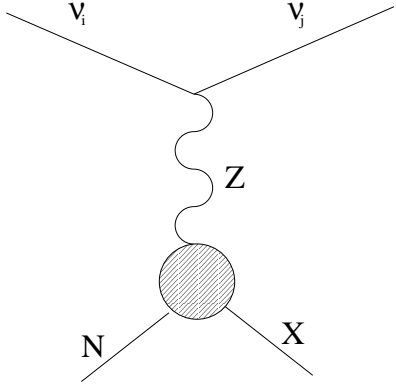, height=2 in}
%angle=270 hscale=50 vscale=50 hoffset=50voffset=-325}
\caption{The $Z$ exchange describes the process $\nu + N \to \nu +X.$
The amplitude responsible for $N-Z-X$ interaction $B_{\nu}$ is schematically
described in the text}
\end{figure}
The amplitudes are given by 
\begin{equation}
A_{i\rightarrow f}^D=\bar{u}_f\gamma ^\mu \left( g_V ^D-g_A^D\gamma _5\right)
u_iD_{\mu \nu }B^\mu , 
\end{equation}
and
\begin{equation}
A_{i\rightarrow f}^M=\left( -g_A^M\right) \left[ \overline{u}_f\gamma ^\mu \gamma_5
u_i-\bar{v}_i\gamma ^\mu  \gamma_5 v_f\right] D_{\mu \nu }B^\nu ,
\end{equation}
where $D_{\mu \nu }$ is the $Z_{0{ \rm  }}$ propagator and $B^\nu $ describes the 
$Z_{0{ \rm  }}$ interaction with nuclei $Z_{0{ \rm  }}N\rightarrow X.$

Both amplitudes look different, but if we approximate them for
relativistic neutrinos $\left( E_\nu >>m_\nu \right) $, with the relation 
\begin{equation}
\overline{v }_i\gamma ^\mu v_f=-\overline{u}_f\gamma ^\mu u_i ,
\end{equation}
and
\begin{equation}
\gamma _5u_i=-u_i+0\left( \frac{m_\nu }{E_\nu }\right) ,
\end{equation}
we find in both cases
\begin{equation}
A_{i\rightarrow f}=\Omega \overline{u}_f\gamma ^\mu u_iD_{\mu \nu }B^\nu ,
\end{equation}
where $\Omega =g_V ^D+g_A^D$ for Dirac and $\Omega =2g_A^M$ for Majorana
neutrinos.

The measurement of the total cross section $\sigma (\nu N\rightarrow \nu X)$
gives one number $\Omega $ and we are not able to say whether $\Omega =g_V
^D+g_A^D$ or $\Omega =2g_A^M$. Therefore, even if the neutral current interaction is
so different for Dirac and Majorana neutrinos, it cannot be used to
distinguish them.

To see possible differences in the behavior of Dirac and Majorana neutrinos,
which could follow from the symmetrization procedure let us consider the process 
\begin{equation}
e^{-}e^{+}\rightarrow \nu _M\nu _M\mbox{   }{ \rm or}\mbox{   } \to 
\nu _D\overline{\nu }_D. 
\end{equation}
We suppose that the measurement of the angular distribution of final neutrinos
(of course in the case that such a distribution is measured which is not the
case up to now) is the simplest way to find their character:
if the angular distribution has forward-backward symmetry the neutrinos are
Majorana particles if not, Dirac neutrinos were produced.

To check whether the  above statement is true, let us calculate the helicity amplitudes
(for simplicity we neglect the electron and neutrino masses)(see for details Ref.
\cite{zr},\cite{gl}). 
For Majorana neutrinos, four helicity amplitudes do not vanish, 
\begin{equation}
M_M(\Delta \sigma =\pm 1,\Delta \lambda =\pm 1)\neq 0, 
\end{equation}
where $\Delta \sigma =\sigma -\overline{\sigma },\Delta \lambda =\lambda -%
\overline{\lambda }$ and $\sigma \left( \overline{\sigma }\right) $ and $%
\lambda (\overline{\lambda })$ are helicities of the electron (positron) and the
final neutrino (antineutrino).

Whereas, there are only two amplitudes for Dirac neutrinos 
\begin{equation}
M_D\left( \Delta \sigma =\pm 1,\Delta \lambda =-1\right) =\sqrt{2}M_M(\Delta
\sigma =\pm 1,\Delta \lambda =-1). 
\end{equation}
If we calculate the unpolarized cross section 
\begin{equation}
\frac{d\sigma }{d\cos \theta }=\frac 14\sum_{\Delta \sigma ,\Delta \lambda }%
\frac{d\sigma (\Delta \sigma ,\Delta \lambda )}{d\cos \theta } ,
\end{equation}
we find out that there is difference between the Dirac and the Majorana cases.
The important feature of a detector, that does not measure helicity is that it
also is not able to distinguish a neutrino from an antineutrino (Fig. 3). Therefore we
have to add the cross section for Dirac neutrinos and antineutrinos. Due to the formulae:
\begin{equation}
\frac{d\sigma ^D}{d\cos \theta }\left( \sigma \right) =\frac{d\sigma ^M}{%
d\cos \theta }\left( \Delta \lambda =-1\right), 
\end{equation}
and
$$
\frac{d\sigma ^D}{d\cos \theta }\left( \pi -\theta \right) =\frac{d\sigma ^M%
}{d\cos \theta }\left( \Delta \lambda =+1\right), 
$$
the final result will be now symmetric.
\begin{figure}
\epsfig{figure=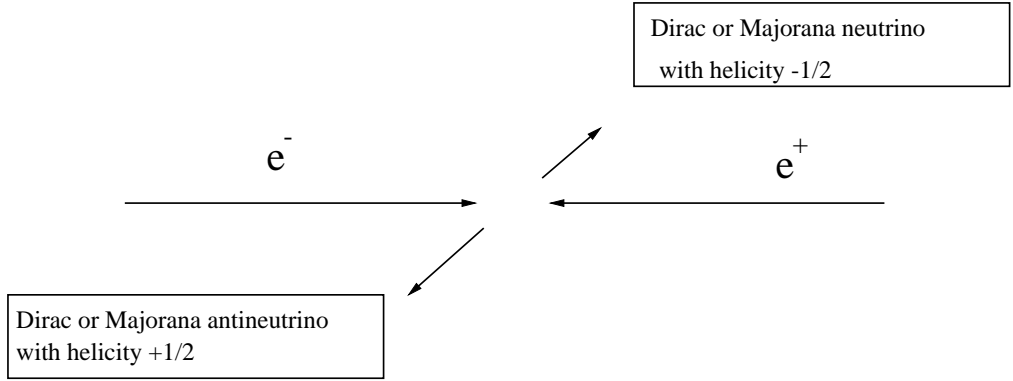, height=2 in}
%angle=270 hscale=50 vscale=50 hoffset=50voffset=-325}
\caption{Detectors do not distinguish lepton number from helicity.}
\end{figure}
For the total cross section we recover once more the equivalence between both
types of neutrinos. In order not to take into account the same spin
configuration two times, we have to integrate the Majorana cross section only
over half of the full solid angle and we have 
\begin{equation}
\sigma _{tot}\left( M\right) =\int_{0}^1d\cos \theta \frac{d\sigma }{d\cos
\theta }=\int_{-1}^1d\cos \theta \frac{d\sigma ^D}{d\cos \theta }=\sigma
_{tot}\left( D\right) . 
\end{equation}
There is only one terrestrial experiment, which currently promises to state
whether neutrinos are Majorana or Dirac particles. It is the neutrinoless
double $\beta $ decay of nuclei $\left( \beta \beta \right) _{0\nu }$ \cite{fur}. 
\begin{equation}
\left( A,Z\right) \rightarrow \left( A,Z+2\right) +2e^{-}. 
\end{equation}
There are many different mechanisms which could be responsible for $\left(
\beta \beta \right) _{0\nu }$ decay \cite{kla}. The most important one is massive
Majorana neutrino exchange \cite{kla} (see Fig. (4)).

It has been proved that independently of the mechanism which governs the $\left(
\beta \beta \right) _{0\nu },$ there is a generic relation between the
amplitude of $\left( \beta \beta \right) _{0\nu }$ decay and the Majorana
mass term for neutrinos \cite{val}. If any of these two quantities vanishes, the
other one vanishes, too, and vice versa if one of them is not zero, the
other also differs from zero.

\begin{figure}
\epsfig{figure=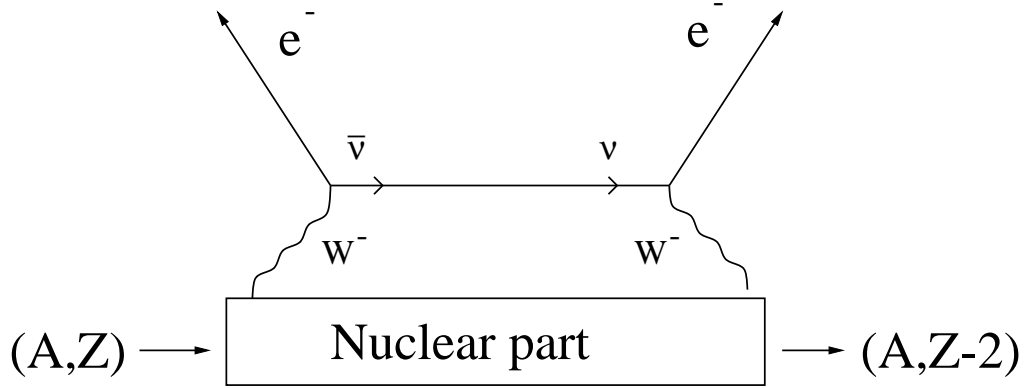, height=2 in}
%angle=270 hscale=50 vscale=50 hoffset=50voffset=-325}
\caption{Massive Majorana neutrino exchange mechanism describing the
 neutrinoless double $\beta$ decay. The antineutrino 
$\bar{\nu}$ emitted in one vertex must be absorbed as 
a neutrino $\nu$ in other. Such a scenario is possible only if 
the neutrino is massive (then there is a chance that the 
 emitted antineutrino has 
negative helicity $\bar{\nu}$ and must be a Majorana particle 
(then $\bar{\nu}=\nu$).}
\end{figure}

Taking into account the most obvious mechanism from Fig.4 the $\left( \beta
\beta \right) _{0\nu }$ amplitude is given by 
\begin{equation}
A_{\left( \beta \beta \right) _{0\nu }}=A_{nucl}\left\langle m_\nu
\right\rangle ,
\end{equation}
where $A_{nucl}$ describes the nuclear transition and $\left\langle m_\nu
\right\rangle $is given by Eq.(5). Many experiments on the search for $%
\left( \beta \beta \right) _{0\nu }$ decay of different nuclei are going on
at present. Unfortunately, up to now such a decay has not been found and
experimentalists can only give a lower bound on the $\left( \beta \beta
\right) _{0\nu }$ decay modes of various nuclei. The most stringent limit was
found in the germanium Heidelberg-Moscow experiment. Their latest result on
the half-life time $T_{1/2}^{0\nu }\sim \left| A_{\left( \beta \beta \right)
_{0\nu }}\right| ^{-2}$ is \cite{eff}
\begin{equation}
T_{1/2}^{0\nu }\left( Ge\right) >5.7 \cdot 10^{25}
\mbox{        }\;{\rm year (\mbox{        } 90\% \; C.L.)} ,
\end{equation}
from which the bound on $\left| \left\langle m_\nu \right\rangle \right| $
(Eq. (5)) has been found. Such results alone give no chance to conclude about
the nature of neutrinos. There are however other experimental data on
mixing matrix elements $U_{ei}$ and masses which are independent of the
neutrino character. This information comes from flavor oscillation
experiments (see Appendix) tritium $\beta $ decay and cosmology. We can use
this data and check whether the bound (Eq. (5) is satisfied. If it is, the
results are still not conclusive. If however the $U_{ei}$ and $m_i$ are such that
the value of $\left\langle m_\nu \right\rangle $ is greater than the present
bound, neutrinos must have Dirac character.

\section{Checking the agreement of $\left( \beta \beta \right) _{0\nu }$
decay bounds with other experimental results.}

The discussion which follows depends on the number of light neutrinos. Three
such neutrinos are necessary to explain solar \cite{sol} and atmospheric \cite{sup}
anomalies. Four light neutrinos must be introduced if, in addition, the LSND
results \cite{lsn} is not disregarded. Here we will present results for three
light neutrinos \cite{cza}. So that we have a relation between 3 flavor states $(\nu
_e,\nu _\mu ,\nu _\tau )$ and 3 eigenmass states $(\nu _1,\nu _2,\nu _3)$%
\begin{equation}
\left( 
\begin{array}{c}
\nu _e \\ 
\nu _\mu \\ 
\nu _\tau 
\end{array}
\right) =\left( 
\begin{array}{ccc}
U_{e1} & U_{e2} & U_{e3} \\ 
U_{\mu 1} & U_{\mu 2} & U_{\mu 3} \\ 
U_{\tau 1} & U_{\tau 2} & U_{\tau 3} 
\end{array}
\right) \left( 
\begin{array}{c}
\nu _1 \\ 
\nu _2 \\ 
\nu _3 
\end{array}
\right) . 
\end{equation}
The three elements in the first row of the mixing matrix $\left( 
\begin{array}{ccc}
U_{e1}, & U_{e2}, & U_{e3} 
\end{array}
\right) $ are the scenario of our discussion.

Besides the $\left( \beta \beta \right) _{0\nu }$ decay there are three main
sources where information about mixing matrix elements $U_{ei}$ and $m_i$
masses of neutrinos are given:

(i) tritium $\beta $ decay

(ii) cosmology (dark matter and number of neutrino species induced by
nucleosynthesis) and most importantly

(iii) solar and atmospheric neutrino oscillation.

Without going into details, we present only the required results (see \cite{cza} for
detail).

TRITUM $\beta $ DECAY.

The latest result from the Curie plot endpoint of tritium $\beta $ decay gives the bound
on 
\begin{equation}
m\left( \nu _e\right) =\left[ \left| U_{e1}\right| ^2m_1^2+\left|
U_{e2}\right| ^2m_2^2+\left| U_{e3}\right| ^2m_3^2\right] ^{1/2}\leq m_\beta ,
\end{equation}
where%
$$
m_\beta = 2.7\;{\rm eV}\; \cite{tro}\;\;\;3.4\;{\rm eV} 
\;\cite{mai} . 
$$
Similar limits on $m\left( \nu _\mu \right) $ and $m\left( \nu _\tau \right) $
 are much larger and less precise, so they are not interesting for our
next analysis.

COSMOLOGY

In order not to exceed the critical density of the Universe the sum of
masses of light, stable neutrinos \cite{un} 
\begin{equation}
\sum_\nu m_\nu <30\;{ \rm  }eV. 
\end{equation}
Then there is no place for cold matter. If only 20\% of all dark matter is
formed by neutrinos then \cite{un}
\begin{equation}
\sum_\nu m_\nu \sim 6\;{ \rm  }eV. 
\end{equation}
Presently the best fit to cosmological observations is obtained if only
30\% of the critical density is formed by dark matter. The rest ($\sim $70\%) is 
explained by the cosmological constant. Then, if all hot dark
matter (20\% of all dark matter) is formed by neutrinos \cite{un}
\begin{equation}
\sum_\nu m_\nu \sim 2\;{ \rm  }eV. 
\end{equation}
There is also a bound on the equivalent number of neutrino species $N_{\nu 
{ \rm  }}$which follows from the present abundance of $^4$He. It was found \cite{li}
that $N_{\nu { \rm  }}\sim (2 \div 4)$ with 95\% C.L.

REACTOR, ATMOSPHERIC and
SOLAR NEUTRINO OSCILLATION \\
From CHOOZ \cite{ch} and the atmospheric neutrino anomaly \cite{sup}  we can find 
\begin{equation}
\left| U_{e3}\right| ^2\leq 0.05 .
\end{equation}
There are three still accepted solutions of solar neutrino deficit \cite{sol} (i)
vacuum oscillation $VO$, (ii) small mixing angle MSW transition
$(SMA)$, and (iii) large mixing angle MSW transition $(LMA)$.

(i)for $VO$ the constraints on $\left| U_{e2}\right| ^2$ are not unique
and two ranges of values are possible (which we denote as small = $S$ or large
= $L$) 
\begin{equation}
0.24\leq \left| U_{e2}^2\right| _S^{(VO)}\leq 0.48, 
\end{equation}
or 
\begin{equation}
0.48\leq \left| U_{e2}^2\right| _L^{VO}\leq 0.76. 
\end{equation}
For the MSW solution it is necessary that $\left| U_{e2}\right| ^2 < \left|
U_{e1}\right| ^2$ in order to fulfill the resonance condition so we have
only one range of values.

(ii) For $SMA$ MSW transition we get: 
\begin{equation}
0.0005\leq \left| U_{e2}^2\right| ^{(SMA)}\leq 0.0026. 
\end{equation}

(iii) Finally for $LMA$ MSW resolution of solar neutrino anomaly there is: 
\begin{equation}
0.204\leq \left| U_{e2}^2\right| ^{(LMA)}\leq 0.48. 
\end{equation}
There are two possible mass schemes, which can describes oscillation data.
They are presented in Fig. 5.

\begin{figure}
\epsfig{figure=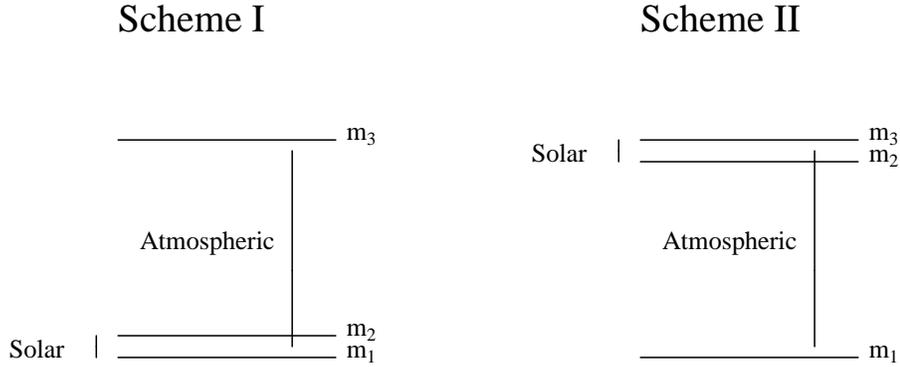, height=2 in}
%angle=270 hscale=50 vscale=50 hoffset=50voffset=-325}
\caption{Two possible neutrino mass spectra which can describe the 
oscillation data.}
\end{figure}

In addition the total scale for neutrino masses is not fixed and different
scenarios are possible (Fig. 6, Eqs. 38, 39, 40)
\begin{figure}
\epsfig{figure=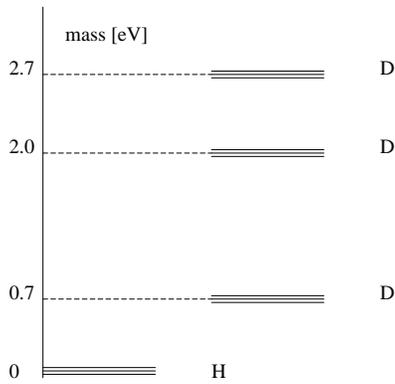, height=2 in}
%angle=270 hscale=50 vscale=50 hoffset=50voffset=-325}
\caption{Different scales for neutrino masses. In the first hierarchical 
scheme (H) 
$m_1 \simeq 0$ and $m_1 << m_2 << m_3$ neutrino masses are too small 
to be responsible for hot dark matter. In all other schemes - almost 
degenerate (D) - neutrinos can explain the existence of the hot dark matter 
without or with a non-vanishing cosmological constant.}
\end{figure}
Now we can combine all the information and check whether the bound on $\left\langle
m_\nu \right\rangle $ (Eq. (5)) is satisfied. In the $\left\langle m_\nu
\right\rangle $ there are squares of $U_{ei}$'s and large
cancellations are possible. From other experiments we have only information
about the modulus, not about phases. If we also take into account, that the scale of
masses is not known the method is not powerful enough. Despite of that, in some cases
the results can be conclusive. For example, for almost degenerate neutrinos if
we know that two elements of mixing matrix are small, then the third must be
large, close to 1. In this case, independently of the possible cancellations, $%
\left\langle m_\nu \right\rangle $ is large ($\left| \left\langle m_\nu
\right\rangle \right| \sim m_1$ and can be greater than the present bound on $%
\left( \beta \beta \right) _{0\nu }$ decay. Then we conclude that neutrinos
must be Dirac particles. For details and discussions of different scenarios
we refer to \cite{cza}.

\section{Conclusions.}

As (i) the SM works very well and no signal about non standard neutrino
interaction is seen, and

(ii) in any of the astrophysical sources and terrestrial experiments neutrinos are
produced with an energy much larger than their mass,

it is extremely difficult to find an experimental signal which would inform us about
the nature of neutrinos.

There is only one terrestrial experimental test that can reveal the Majorana
character of neutrinos - the neutrinoless double $\beta $ decay.
Unfortunately, experimental groups placed only the upper limit on the $%
\left( \beta \beta \right) _{0\nu }$ decay half life time. If neutrinos are
Majorana particles, probably the next experiments which measure $%
\left\langle m_\nu \right\rangle $ up to 0.01 eV or 0.001 eV have a chance to
measure it.

If neutrinos are Dirac particles we should get a signal about it by
confronting the $\left( \beta \beta \right) _{0\nu }$ bound with independent
information about masses and mixing matrix elements.

The present experimental precision is not good enough to find the answer.
However, we are able to get some partial information e.g.
if $SMA$ MSW mechanism describes the solar neutrino deficit, and almost degenerate
neutrinos have $m_\nu >0.22$ eV then they must be Dirac particles. If the
future GEMINI experiment still gives only a bound on $\left\langle m_\nu
\right\rangle $ the next solar neutrino measurements (SNO and BOREXINO) have a
chance to state that neutrinos are Dirac particles.

\section{Appendix}

We would like to clarify what the  formulae for flavor oscillation $P_{\alpha
\rightarrow \beta }(x)$ and effective neutrino mass $\left\langle m_\nu
\right\rangle $ look like for Dirac neutrinos.

The mass term of n Dirac neutrinos is%
\begin{equation}			
L_{mass}=-\overline{\nu }_LM_D\nu _R+h.c.=-\frac 12\left( \overline{\nu }_L,
 \;\; \overline{\nu }_L^c\right) M_\nu \left( {\nu _R}, \;{\nu _R^c}\right)^T
+h.c., 
\end{equation}
where $M_D$ is an arbitrary $n \times n$ matrix,%
\begin{equation}
\nu _{R(L)}^c=c\overline{\nu }_{L(R)}^T,\;\;{\rm and }\;\;M_\nu =\left( 
\begin{array}{cc}
0 & M_D \\ 
M_D^T & 0 
\end{array}
\right) . 
\end{equation}
The $M_D$ matrix can be diagonalized by the biunitary transformation%
\begin{equation}
M_D\rightarrow V^TM_DV^{\prime }=\left( M_D\right) _{diag}, 
\end{equation}
where $V$ and $V^\prime $ are the $n \times n$ unitary matrices.

Then the $M_\nu $ matrix is diagonalized by the transformation%
\begin{equation}
M_\nu \rightarrow U^TM_\nu U=\left( 
\begin{array}{cc}
\left( M_D\right) _{diag} & 0 \\ 
0 & \left( M_D\right) _{diag} 
\end{array}
\right) ,
\end{equation}

where the $2n \times 2n$ matrix U is%
\begin{equation}
U=\frac 1{\sqrt{2}}\left( 
\begin{array}{cc}
-iV, & V \\ 
iV^{\prime }, & V^{\prime } 
\end{array}
\right) . 
\end{equation}
In the mass eigenstate basis for charged leptons the left-handed charged
current interaction 
\begin{equation}
L_{CC}\equiv -\frac g{\sqrt{2}}\overline{\nu }_L\gamma ^\mu l_LW_\mu
^{+}+h.c 
\end{equation}
can be written in the form%
\begin{equation}
L_{CC}\equiv -\frac g{\sqrt{2}}\overline{\Psi }_LV^T\gamma ^\mu l_LW_\mu
^{+}+h.c ,
\end{equation}
where%
$$
\Psi _L=\frac 1{\sqrt{2}}\left( iN_{1L}+N_{2L}\right) ,
$$
and two Majorana bispinors $N_{1i}$ and $N_{2i}$ correspond to the same mass
eigenvalue $m_i$ of the matrix (49)%
\begin{equation}
\nu _L=\frac 1{\sqrt{2}}\left( iV^{*},V^{*}\right) \left( 
\begin{array}{c}
N_{1L} \\ 
N_{2L} 
\end{array}
\right) =V^* {\Psi _L}. 
\end{equation}
Now the effective neutrino mass is %
\begin{equation}
\left\langle m_\nu \right\rangle
=\sum_{i=1}^{2n}U_{ei}^2m_i=\sum_{l=1}^n\frac 12\left( \left(
-iV_{ei}\right) ^2+\left( V_{ei}\right) ^2\right) m_i=0 . 
\end{equation}
For the probability that  a neutrino born with flavor $\alpha $ will have
flavor $\beta $ after traveling distance $x$ we get%
\begin{eqnarray}
P_{\alpha \rightarrow \beta }(x)&=&
\left| \sum\limits_{i=1}^{2n}U_{\beta
i}^{*}U_{\alpha i}e^{-i 
\frac{m_i^2}{2p}x}\right| ^2 \nonumber \\
&=&\left| \sum\limits_{i=1}^n e^{-i\frac{m_i^2x}{2p}}\frac
12\left\{ \left( iV_{\beta i}^{*}\right) \left( -iV_{\alpha i}\right)
+\left( V_{\beta i}^{*}\right) \left( V_{\alpha i}\right) \right\} \right|
^2 \nonumber \\
&=& 
\left| \sum\limits_{i=1}^ne^{-i\frac{m_i^2x}{2p}}V_{\beta i}^{*}
V_{\alpha i}^{*} \right| ^2 .
\end{eqnarray}

We see that $P_{\alpha \rightarrow \beta }(x)$ looks exactly the same for $n$
Dirac and n Majorana neutrino oscillation. The only difference is the number
of CP violating phases in mixing matrices $V_{\alpha i}$ in both cases. They 
are $\left( n-1\right) \left( n-2\right) /2$ for Dirac neutrinos and
$n(n-1)/2$ for Majorana neutrino mixing.
However, the physical phases by which the mixing matrices differ do not
enter into transition probabilities \cite{a1}. Consequently by studying neutrino
oscillation in vacuum or in matter it is impossible to distinguish the
nature of neutrinos \cite{a2}.

\end{document}